\documentclass[reprint,prl,aps]{revtex4-1}

\usepackage{color}
\usepackage{fancyhdr}
\usepackage{graphicx}
\usepackage[bookmarks]{hyperref}
\usepackage{emptypage}
\usepackage{natbib,comment,bm, amsmath, wasysym, color, feynmf}
\begin{document}

\title{Large Tensor Non-Gaussianity from Axion-Gauge Fields Dynamics}
\author{Aniket Agrawal}
\affiliation{Max Planck Institute for Astrophysics,
Karl-Schwarzschild-Str. 1, 85748 Garching, Germany}
\author{Tomohiro Fujita}
\affiliation{Stanford Institute for Theoretical Physics and Department
of Physics, Stanford University, Stanford, CA 94306, U.S.A.}
\affiliation{Department of Physics, Kyoto University, Kyoto 606-8502, Japan}
\author{Eiichiro Komatsu}
\affiliation{Max Planck Institute for Astrophysics,
Karl-Schwarzschild-Str. 1, 85748 Garching, Germany}
\affiliation{Kavli Institute for the Physics and Mathematics of the Universe (Kavli IPMU, WPI), Todai Institutes for Advanced Study, the University of Tokyo, Kashiwa 277-8583, Japan}
\date{\today}
\begin{abstract}
We show that an inflation model in which a spectator axion field
 is coupled to an SU(2) gauge field produces a large three-point
 function (bispectrum) of primordial gravitational waves, $B_{h}$, on
 the scales relevant to the cosmic microwave background experiments.
 The amplitude of the bispectrum at the equilateral configuration is 
 characterized by $B_{h}/P_h^2=\mathcal{O}(10)\times \Omega_A^{-1}$,
 where $\Omega_A$ is a fraction of the energy density in the gauge field
 and $P_h$ is the power spectrum of gravitational waves produced by the
 gauge field.
\end{abstract}
\maketitle

\section{I. introduction}
Quantum vacuum fluctuations in quasi-de Sitter space in the early
universe (cosmic inflation
\cite{sato:1981,guth:1981,linde:1982,albrecht/steinhardt:1982}) produce
a stochastic background of tensor metric perturbations (gravitational
waves; GWs) \cite{grishchuk:1974,starobinsky:1979,abbott/wise:1984},
which creates temperature anisotropies
\cite{rubakov/sazhin/veryaskin:1982,fabbri/pollock:1983,starobinsky:1985}
and polarization
\cite{polnarev:1985,crittenden/davis/steinhardt:1993,seljak/zaldarriaga:1996,kamionkowski/kosowsky/stebbins:1996}
of the cosmic microwave background (CMB).

The power spectrum of GWs from vacuum fluctuations (i.e., the homogeneous solution to the wave equation of GW) is proportional to
the energy scale of inflation. However, this relationship does not hold
if GWs are produced by other sources, e.g., an inhomogeneous solution sourced by scalar fields
\cite{cook/sorbo:2012,carney/etal:2012,biagetti/fasiello/riotto:2013,senatore/silverstein/zaldarriaga:2014},
a U(1) gauge field
\cite{sorbo:2011,anber/sorbo:2012,barnaby/peloso:2011,barnaby/etal:2012,peloso/sorbo/unal:2016}, a non-Abelian
SU(2) gauge field
\cite{maleknejad/sheikh-jabbari:2013,dimastrogiovanni/peloso:2012,adshead/etal:2013,adshead/martinec/wyman:2013,maleknejad:2016,dimastrogiovanni/fasiello/fujita:2016},
etc. These sourced GWs are typically non-Gaussian, yielding a non-zero
three-point function (bispectrum) of tensor metric perturbations.

The model we study in the paper contains three fields: inflaton, a
spectator pseudo-scalar field, and a gauge field. The latter two fields
are coupled, whereas the inflaton field is coupled only
gravitationally. Cook and Sorbo \cite{cook/sorbo:2013} calculated the
bispectrum of GWs from a U(1) field, finding a large value.
However, the
amplified U(1) field produces perturbations in the inflaton field which,
in turn, produce the scalar curvature perturbation that is also
non-Gaussian \cite{ferreira/sloth:2014}. Avoiding large contributions
to the scalar power spectrum and bispectrum that are incompatible with
the observational data puts severe restrictions on
the model \cite{namba/etal:2015}: the GWs cannot be produced over a wide
range in wavenumbers but have to be localized.

Here, we calculate the bispectrum of GWs sourced by an SU(2)
field \cite{dimastrogiovanni/fasiello/fujita:2016}, finding a large
value. Unlike for the U(1) model, the tensor component of the SU(2)
field is amplified, but the scalar components are not amplified in the
relevant parameter space, and thus the sourced scalar curvature
perturbation remains small compared to the vacuum contribution, allowing
for production of significant GWs over a wide range in
wavenumbers. Most importantly, GWs are produced {\it
linearly} by the tensor component of the SU(2) field, whereas in the
U(1) model they are produced non-linearly by a product of the
fields. Thus, the bispectrum is produced by the tree-level diagrams in
this model, rather than by loop diagrams as in the U(1) model.

\section{II. Model}
The Lagrangian density of the model is given
by~\cite{dimastrogiovanni/fasiello/fujita:2016}
\begin{equation}
 \mathcal{L}=\mathcal{L}_{GR}+\mathcal{L}_{\phi}+\mathcal{L}_{\chi}-\frac{1}{4}F_{\mu\nu}^{a}F^{a\mu\nu}+\frac{\lambda\,\chi}{4f}F_{\mu\nu}^{a}\tilde{F}^{a\mu\nu}\,,
\label{model action}
\end{equation}
where $\mathcal{L}_{GR}$, $\mathcal{L}_{\phi}$, and $\mathcal{L}_{\chi}$
are the Lagrangian densities of the Einstein-Hilbert action and the
canonical actions for an inflaton field $\phi$ and a pseudo-scalar
``axion'' field $\chi$, respectively. Repeated indices are
summed. $\lambda$ and $f$ are dimensionless and dimensionful constants,
respectively. The field strength of the SU(2) field, $A_\nu^a$, is given
by
$F_{\mu\nu}^a\equiv\partial_{\mu}A^a_{\nu}-\partial_{\nu}A^a_{\mu}-g\epsilon^{abc}A^b_{\mu}A^c_{\nu}$,
and $\tilde{F}^{a}_{\mu\nu}$ is its dual. $\epsilon^{abc}$ is the
anti-symmetric Levi-Civita symbol and $g$ is a dimensionless
self-coupling constant. This action was inspired by the chromo-natural
inflation model~\cite{adshead/wyman:2012} in which there was no $\phi$
but $\chi$ played the role of inflaton.

\begin{figure*}
        \centering
        \includegraphics[width=1\textwidth]{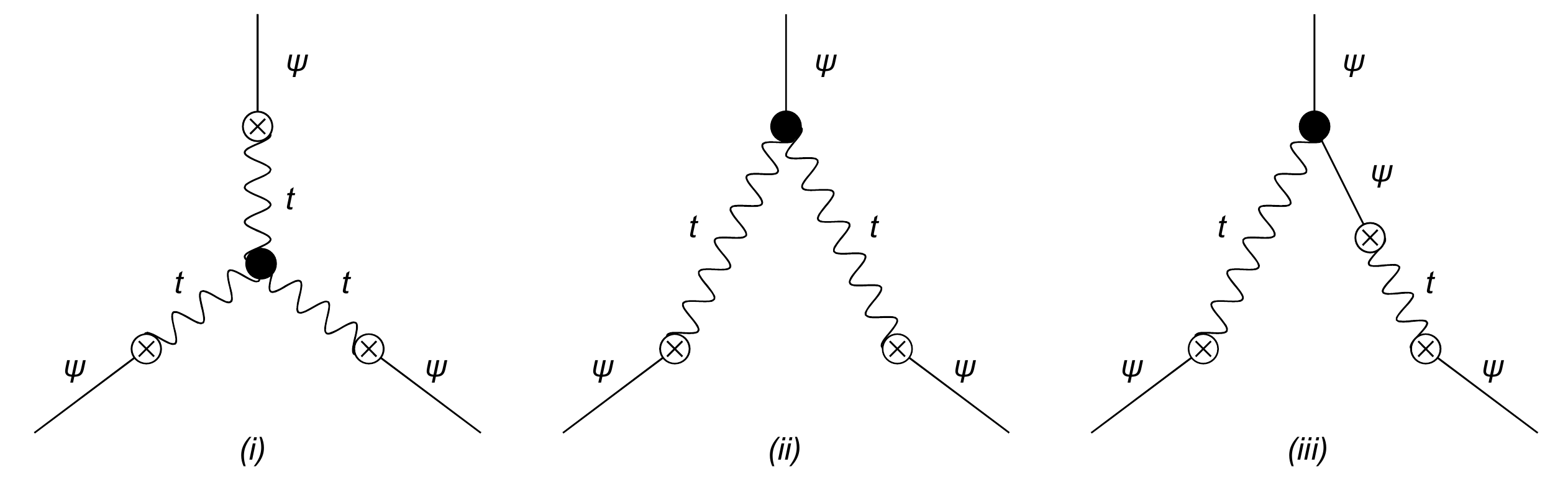}
       \caption{Feynman diagrams illustrating the tree-level contributions
 from the cubic interactions $L_3^{(i)}, L_3^{(ii)}$ and $L_3^{(iii)}$
 to the bispectrum of GWs. The straight and wavy lines show $\psi_{ij}$
 and $t_{ij}$, respectively. The black dots show the vertices of the
 three-point interactions, while the circled crosses show the mixing
 between $\psi_{ij}$ and $t_{ij}$ (the third line in
 Eq.~\eqref{quadratic L}). }
\label{Feynman}
\end{figure*}

At the background level, the axion and the gauge fields have the
slow-roll attractor solution, $A^a_0 = 0$ and $A^a_i = aQ\delta^a_i$,
where $a(t)$ in the pre-factor (not the index) is the scale factor and $Q\equiv (-f \partial_\chi
U/3g\lambda H)^{1/3}$ with $U$ being the potential of $\chi$ and
$H\equiv d\ln a/dt$ the expansion rate during inflation. It is always
possible to keep this configuration against a spatial rotation by
performing the corresponding $SU(2)$ gauge transformation, because the
SU(2) gauge group is isomorphic to SO(3)~\cite{obata/soda:2016}. Then we
identify the gauge index $a$ with a spatial index. The
transverse-traceless part of the perturbation of $A^a_i$ behaves as a
tensor perturbation $t_{ai}$:
\begin{equation}
\delta A^a_i = t_{ai}+\cdots,\quad 
t_{ii}= \partial_a t_{ai} = \partial_i t_{ai}= 0,
\end{equation}
where $\cdots$ denotes scalar and vector components
\cite{dimastrogiovanni/peloso:2012}. 
The tensor metric perturbation, $h_{ij}$, is defined as
$ds^2=-dt^2+a^2(t)(\delta_{ij}+h_{ij})dx^idx^j$ with
$h_{ii}=\partial_j h_{ij}=0$.   
In this paper, we mainly use a canonically normalized field
$\psi_{ij} \equiv aM_{\text{Pl}}h_{ij}/2$ instead of $h_{ij}$.
The axion field obeys the slow-roll equation
\begin{align}
\dot{\chi} = 2fH \frac{m_Q+m_Q^{-1}}{\lambda}\,,
\label{VEV}
\end{align}
with $m_Q\equiv gQ/H$.

The potential of $\chi$ determines time evolution of $m_Q$. In this
paper we do not specify the form of the potential but assume that $m_Q$ is a
constant. This gives scale-invariant tensor perturbations sourced by the gauge field.

The sufficient condition for
the absence of instability in the scalar components of the
SU(2) field is $m_Q> \sqrt{2}$ \cite{dimastrogiovanni/peloso:2012}. On the other
hand, we cannot ignore back-reaction of $t^2$ on the background solution
when $m_Q$ is too large. The relevant parameter space depends on the
expansion rate $H$ during inflation, and it is $m_Q=$~a few for the
tensor-to-scalar ratio of the vacuum contribution of order
$10^{-5}-10^{-2}$. See Ref.~\cite{fujita/namba/tada:prep} for low-scale
inflation allowing for a large $m_Q$.

Expanding Eq.~\eqref{model action}
with respect to $\psi_{ij}$ and $t_{ij}$, we obtain
the following perturbative Lagrangian at the quadratic and cubic order as
$\mathcal{L}_{\rm tensor}=a^{-3}[L_2 + L_3^{(i)} + L_3^{(ii)} +L_3^{(iii)}]$, 
with
\begin{multline}
L_2=\frac{1}{2}\psi'_{ij}\psi'_{ij}-\frac{1}{2}\partial_k\psi_{ij}\partial_k \psi_{ij}+\frac{1}{\tau^2}\psi_{ij}\psi_{ij}+\frac{1}{2}t'_{ij}t'_{ij} 
\\
-\frac{1}{2}\partial_l t_{ij}\partial_l t_{ij}
+\frac{2m_Q+m_{Q}^{-1}}{\tau}\epsilon^{ijk} t_{il}\partial_j t_{kl}-\frac{m_Q ^2+1}{\tau^2}t_{ij}t_{ij}
\\
+\frac{2\sqrt{\epsilon_B}}{\tau}\left[\frac{1}{m_Q}\psi_{ij}t'_{ij}-\psi_{jm}\epsilon_{aij}\partial_i t_{am} +\frac{m_Q}{\tau}\psi_{ij}t_{ij}\right],
\label{quadratic L}
\end{multline}
\begin{multline}
L_3^{(i)}=c^{(i)}\Bigg[\epsilon^{abc}t_{ai}t_{bj}\left(\partial_i t_{cj}-\frac{m_Q^2+1}{3m_Q\tau}\epsilon^{ijk}t_{ck}\right)
\\-\frac{ m_Q}{\tau} t_{ij}t_{jl}t_{li}\Bigg],
\end{multline}
\begin{multline}
L_3^{(ii)}=c^{(ii)}\psi_{ij} \bigg[\frac{\tau}{2m_Q}\Big\{ t_{il}' t_{jl}'- \partial_i t_{kl} (\partial_j t_{kl}-2 \partial_k t_{jl})
\\\qquad\qquad\qquad-\partial_k t_{il} \partial_k t_{jl} \Big\} -\epsilon^{iab}t_{al}\left( \partial_j t_{bl}-\partial_l t_{bj}\right)
\\
\left.    -\epsilon^{lab} t_{ai}\partial_l t_{bj}
- \frac{3m_Q}{2\tau} t_{il} t_{jl} \right],
\end{multline}
\begin{multline}
L_3^{(iii)}=c^{(iii)}\psi_{ij}\bigg[\frac{1}{m_Q}
\psi_{jk} t_{ik}'
\\
+\epsilon^{ajm}\psi_{lm}\partial_i t_{al}
-\psi_{jk} \epsilon^{akl} \partial_l t_{ai}\bigg],
\end{multline}
where $\tau\simeq -1/aH$ is the conformal time, prime denotes 
derivative with respect to $\tau$ and  we neglect the
$\mathcal{O}(\psi^3)$ terms and terms suppressed by slow-roll
parameters. We organize terms such that $L_3^{(i)}=\mathcal{O}(t^3),
L_3^{(ii)}=\mathcal{O}(\psi t^2)$ and $L_3^{(iii)}=\mathcal{O}(\psi^2
t)$. Their tree-level contributions to the tensor bispectrum are
illustrated as Feynman diagrams in Fig.~\ref{Feynman}.

The coefficients of the cubic Lagrangians are $c^{(i)}=g=m_Q^2
H/\sqrt{\epsilon_B}M_{\rm Pl},\  c^{(ii)}=2m_QH/M_{\rm Pl}$, and
$c^{(iii)}=4\sqrt{\epsilon_B}H/M_{\rm Pl}$. Their sizes are
hierarchical, obeying 
\begin{equation}
\frac{c^{(ii)}}{c^{(i)}}=\frac{c^{(iii)}}{c^{(ii)}}=\frac{2\sqrt{\epsilon_B}}{m_Q} \ll 1\,.
\end{equation}
The parameter $\epsilon_B$ is comparable to the energy density fraction
of the gauge field $\Omega_A\equiv \rho_A/\rho_{\rm total}$; thus, it is
tiny in our spectator model,
\begin{equation}
\epsilon_B \equiv \frac{g^2Q^4}{H^2M^2_{\rm Pl}} \simeq \frac{2\Omega_A}{1+m_Q^{-2}}
\ll 1.
\end{equation}

As we find that the contribution from $L_3^{(iii)}$ is negligible
compared to the other two, we focus on $L_3^{(i)}$ and $L_3^{(ii)}$
hereafter.

\section{III. calculation of the Bispectrum}

To solve the dynamics of $\psi_{ij}$ and $t_{ij}$, it is useful to
decompose them with the circular polarization tensors,
\begin{align}
X_{ij}(\tau, \bm{x}) &= \int \frac{{\rm d}^3k}{(2\pi)^3}e^{i\bm k\cdot\bm x}\Big[e^R_{ij}({\bm k})X_{\bm k}^R(\tau)+e^L_{ij}({\bm k})X_{\bm k}^L(\tau)\Big],
\end{align}
where $X=\psi, t$ and the polarization tensors satisfy
$e^{L}_{ij} (-{\bm{k}})= e^{L*}_{ij} ({\bm{k}})=e^{R}_{ij}
({\bm{k}})$ and $i \epsilon_{ijk} k_i e_{jl}^{p}({\bm{k}})=
\pm k e_{kl}^{p}({\bm{k}})$ with $p=R,L$. We normalize $e^p_{ij}$
such that $e_{ij}^R(\bm{k}) e_{ij}^R(-\bm{k})=e_{ij}^L(\bm{k})
e_{ij}^L(-\bm{k})=1$.

Now we quantize the fields and expand them in a perturbative
series~\cite{seery:2008},
\begin{equation}
\hat{X}^{p}_{\bm k}(\tau) = \hat{X}_1^{p}(\tau, \bm k)+\hat{X}_2^{p}(\tau, \bm k)+\ldots
\end{equation}
The first order components are written as
\begin{align}
\hat{t}_1^{p}(\tau, \bm k)&=T_1^{p}(\tau, k)\, \hat{a}^p_{\bm k}+T^{p*}_1(\tau, k)\, \hat{a}^{p\dag}_{-\bm k},
\\
\hat{\psi}_1^{p}(\tau, \bm k)&=\Psi_1^{p}(\tau, k)\, \hat{a}^p_{\bm k}+\Psi^{p*}_1(\tau, k)\, \hat{a}^{p\dag}_{-\bm k},
\end{align}
with the creation/annihilation operators, $\hat{a}^p_{\bm k}$ and
$\hat{a}^{p\dag}_{\bm k}$, satisfying $[\hat{a}_{\bm k}^p, \hat{a}_{-\bm
k'}^{q\dag}]=(2\pi)^3\delta^{pq}\delta(\bm k+\bm k')$. We only consider
 GWs sourced by the gauge field in this paper, and assign
$\hat{\psi}_1$ with the same quantum operator as $\hat{t}_1$. The mode
functions of $\hat{X}_1^p$ satisfy linearized equations of motion
and their solutions induce the second order fields $\hat{X}_2^p$ through
non-linear terms in the equations of motion.

In Ref.~\cite{dimastrogiovanni/fasiello/fujita:2016}, the linearized
equations have been solved. One of the two polarization modes, $T_1^{R}$
or $T_1^{L}$, is amplified, as the background pseudo-scalar $\chi$
spontaneously breaks parity symmetry. Without loss of generality, we
assume that the right-handed mode $T_1^R$ is amplified and ignore the
left-handed mode $T_1^L$. Then its homogeneous solution is given by
\begin{equation}\label{eq:tk_homo}
T_1^R (\tau, k)= \frac{1}{\sqrt{2k}}e^{\frac{\pi}{2} (2m_Q+m_Q^{-1})} W_{\beta, \alpha}(2ik\tau),
\end{equation}
where $W_{\beta, \alpha}(x)$ is the Whittaker function, $\alpha \equiv
-i \sqrt{2m_Q^2+2-1/4}$ and $\beta \equiv -i (2m_Q+m_Q^{-1})$. Using the 
Green's function for $\psi$,
\begin{multline}
G_{\psi}(k,\tau,\eta) = \frac{\Theta(\tau-\eta)}{k^3\tau\eta}\Big[k(\eta-\tau)\cos(k(\tau-\eta))
\\+(1+k^2\tau\eta)\sin(k(\tau-\eta))\Big],
\end{multline}
the sourced GW at first order is obtained as
\begin{equation}
\Psi_1^R (\tau, k)=  \int^{\infty}_{-\infty} {\rm d} \eta\, G_\psi(k,\tau,\eta) D_k(\eta) T_1^R(\eta,k),
\end{equation}
where $D_k(\eta)\equiv \frac{2\sqrt{\epsilon_B}}{m_Q \eta}\partial_\eta
+ \frac{2\sqrt{\epsilon_B}}{\eta^2} \big(m_Q+k\eta\big)$ and $\Theta(x)$
is the Heaviside function. This integration can be done analytically and
the resultant tensor power spectrum in the super horizon limit is
\begin{equation}
 \frac{k^3}{2\pi^2}P_h^{\rm sourced}= \frac{\epsilon_B H^2}{\pi^2 M_{\rm
  Pl}^2}\left|\mathcal{F}(m_Q)\right|^2,
  \label{eq:Ph}
\end{equation}
where the power spectrum is defined by $\langle \hat h_R({\bm k})\hat
h_R({\bm k}')\rangle=(2\pi)^3\delta(\bm{k}+\bm{k}')P_h(k)$. We ignore
the contribution from the left-handed mode. The function $\mathcal{F}$
is given approximately by 
$|\mathcal{F}(m_Q)|\approx e^{2.06m_Q-0.12}$ for $3\le m_Q\le 4$.
The exact expression can be found in
Ref.~\cite{dimastrogiovanni/fasiello/fujita:2016}. Note that $\mathcal{F}(m_Q)$
here is $\mathcal{F}_B+\mathcal{F}_E/m_Q$ there.

The second order $\hat{\psi}^R_{\bm k}$ is induced by $L_3^{(i)}$ through $\hat{t}_2^R$ and by $L_3^{(ii)}$ through $\mathcal{O}(\hat{t}_1^R\times\hat{t}_1^R)$ terms in Fourier space,
\begin{multline}
\hat{\psi}_2^R(\tau,\bm k)= \int^{\infty}_{-\infty} {\rm d} \eta\, G_\psi(k,\tau,\eta) \\ 
\times\left[D_k(\eta) \hat{t}_2^R(\eta, \bm k)+ e^{R*}_{ij}(\hat{\bm k})\int{\rm d}^3x \,e^{-i\bm k\cdot \bm x}\, \frac{\delta L_3^{(ii)}}{\delta\psi_{ij}}\right].
\end{multline}
The second order $\hat{t}^R_{\bm k}$ is given by
\begin{multline}
\hat{t}_2^R(\eta, \bm k) = \int^{\infty}_{-\infty} {\rm d} \eta'\, G_t(k,\eta,\eta')
\, e^{R*}_{ij}(\hat{\bm k})\int{\rm d}^3x \,e^{-i\bm k\cdot \bm x}\, \frac{\delta L_3^{(i)}}{\delta t_{ij}},
\end{multline}
where $G_t(k,\tau,\eta) =
i\Theta(\tau-\eta)[T_1^R(\tau,k)T^{R*}_1(\eta,k)-T^{R*}_1(\tau,k)T_1^R(\eta,k)]$
is the Green's function for $t_2$. $t_{kl}(\tau,\bm x)$ in $\delta L_3^{(i)}/\delta t_{ij}$ and
$\delta L_3^{(ii)}/\delta t_{ij}$ should be evaluated by the first order,
$\int \frac{{\rm d}^3 p}{(2\pi)^3} e^{i\bm p\cdot \bm x}
e^{R}_{kl}(\hat{\bm p})\,\hat{t}_1^R(\tau,\bm p)$. Once $\hat{\psi}_2^R$
is obtained, the tensor three point function is calculated as $\langle
\hat{\psi}_1(\tau,
\bm{k}_1)\hat{\psi}_1(\tau,\bm{k}_2)\hat{\psi}_2(\tau,\bm{k}_3)\rangle $
and its permutations in the leading order.

\section{IV. Results}
We define the bispectrum of the right-handed modes in the super horizon limit as
\begin{equation}
\langle \hat{h}_R(\bm k_1) \hat{h}_R(\bm k_2) \hat{h}_R(\bm k_3)\rangle=(2\pi)^3\delta\left(\sum_{i=1}^3\bm k_i\right)B_{h}^{RRR}(k_1,k_2,k_3).
\end{equation}
We find that the contributions from the diagrams (i) and (ii) in
Fig.~\ref{Feynman} dominate. The contribution from $ L_3^{(i)}$ is
\begin{multline}
k^2_1k^2_2k^2_3 B_h^{(i)}(k_1, k_2, k_3) = 8m_Q^2\Xi \epsilon_Be^{2\pi(2m_Q+m_Q^{-1})}(H/M_{\rm Pl})^4
\\\times\Big[\mathcal{F}^{*2}\mathcal{N}_1+ r^{-2}_2|\mathcal{F}|^2\mathcal{N}_2 +r^{-2}_3\mathcal{F}^2\mathcal{N}_3\Big],
\end{multline}
with $r_i=k_i/k_1\ (i=1,2,3)$. The triangle condition demands
$|r_i-r_j|\le r_k\le r_i+r_j$; the bispectrum vanishes otherwise. The
other functions are defined as
\begin{equation}
 \label{eq:xi}
\Xi \equiv \frac{(1+r_2+r_3)^3}{64r^2_2r^2_3}(r_2+r_3-1)(1+r_2-r_3)(1+r_3-r_2), 
\end{equation}
and
\begin{align}
&\mathcal{N}_i \equiv \int_0^{x_{\rm max}}\frac{dy}{y^2}\,[r_i y\cos(r_i y)-\sin(r_i y)]
\notag\\&\qquad\times\Big[m^{-1}_Q\partial_y+\Big(m_Qy^{-1}-r_i\Big)\Big]
\notag\\\nonumber&\qquad\times \int_{y}^{x_{\rm max}} dz\, {\rm Im}[W^*_{\beta, \alpha}(-2ir_iy)W_{\beta, \alpha}(-2ir_iz)]
\\&\qquad\times \Big(1+r_2+r_3-\frac{5m_Q+2m_Q^{-1}}{z}\Big)\, \mathcal{W}_i(z),
\end{align}
where   $\mathcal{W}_1(z)=W_{\beta,\alpha}(-2ir_2 z)W_{\beta,\alpha}(-2ir_3z)$,  $\mathcal{W}_2(z)=W^*_{\beta,\alpha}(-2i z)W_{\beta,\alpha}(-2ir_3z)$,
and $\mathcal{W}_3(z)=W^*_{\beta,\alpha}(-2ir_2 z)W^*_{\beta,\alpha}(-2iz)$.
We have introduced the UV cutoff $x_{\max}\equiv
2m_Q+m_Q^{-1}+\sqrt{2m_Q^2+2+m_Q^{-2}}$ to avoid incorporating
unphysical vacuum contributions. The integration result is not sensitive
to the cutoff~\cite{dimastrogiovanni/fasiello/fujita:2016}.

The contribution from $L_3^{(ii)}$ is 
\begin{multline}
k^2_1k^2_2k^2_3 B_h^{(ii)}(k_1, k_2, k_3) = 4\Xi \epsilon_B e^{\pi(2m_Q+m_Q^{-1})}(H/M_{\rm Pl})^4
\\\times\Big[\mathcal{F}^{*2}\tilde{\mathcal{N}}_1+ r^{-1}_2|\mathcal{F}|^2\tilde{\mathcal{N}}_2 +r^{-1}_3\mathcal{F}^2\tilde{\mathcal{N}}_3\Big],
\end{multline}
with
\begin{align}
&\tilde{\mathcal{N}}_i \equiv \int_0^{x_{\rm max}}\frac{dy}{y}\,[r_i y\cos(r_i y)-\sin(r_i y)] \Big[y \tilde{\mathcal{W}}_i(y)
\notag\\&+ \Big(\frac{r_1 r_2 r_3}{r_i}y-(r_1+r_2+r_3-r_i)m_Q+\frac{3m_Q^2}{y}\Big)\, \mathcal{W}_i(y)\Big],
\end{align}
where $\tilde{\mathcal{W}}_1(y)=\partial_y W_{\beta,\alpha}(-2ir_2
y)\partial_y W_{\beta,\alpha}(-2ir_3y)$,
$\tilde{\mathcal{W}}_2(y)=\partial_y W^*_{\beta,\alpha}(-2i y)\partial_y
W_{\beta,\alpha}(-2ir_3y)$, and $\tilde{\mathcal{W}}_3(y)=\partial_y
W^*_{\beta,\alpha}(-2ir_2 y) \partial_yW^*_{\beta,\alpha}(-2iy)$.

\begin{figure}
        \includegraphics[width=80mm, height=50mm]{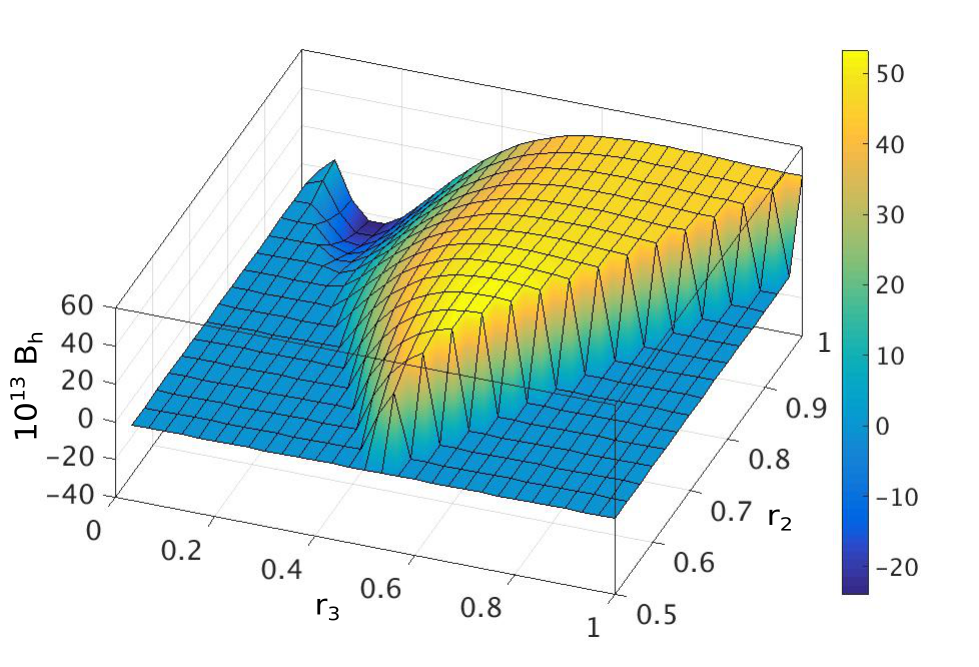}
       \caption{The 3D plot of the numerical result of $10^{13}(k_1
 k_2k_3)^2 (B_{h}^{(i)}+B_{h}^{(ii)})$. Only $r_3\le r_2$ is shown. The
 bispectrum vanishes for $r_2+r_3<1$ by the triangle condition.}
\label{Bhhh}
\end{figure}

In Fig.~\ref{Bhhh}, we plot the bispectrum for $m_Q=3.45$ and
$\epsilon_B=3\times 10^{-5}$, which yield the tensor-to-scalar ratio parameter
of the sourced GW of $r_{\rm sourced}= 0.0472$. The expansion rate
during inflation is $H=1.28\times 10^{13}$~GeV, and the vacuum
contribution (including both right- and left-handed modes) is $r_{\rm
vac}= 0.00256$. We only show $r_3\le r_2$ to avoid duplication.

We find that the bispectrum vanishes in the so-called ``folded limit'',
$r_2+r_3=1$. This appears to be true generally for the bispectrum of
right-handed modes at the tree level. This is a consequence of the
contraction of three polarization tensors. For example, trace of the
product of three polarization tensors, $e^R_{ij}({\bm k}_1)e^R_{jk}({\bm
k}_2)e^R_{ki}({\bm k}_3)$, is equal to $\Xi$ (Eq.~\eqref{eq:xi}), which
vanishes in the folded limit because it contains $r_2+r_3-1$. We find
that other possible contractions of three polarization tensors
multiplying derivative operators and $\epsilon^{ijk}$ are also
proportional to $r_2+r_3-1$ at the tree level.

The shape of the bispectrum is similar to the so-called equilateral
template, $F^{\rm eq}(k_1,k_2,k_3)$ \cite{creminelli/etal:2006}, but is
different in details. When $r_2$ and $r_3$ are comparable it rises
sharply from zero at the folded limit, reaches the maximum at
$r_2=r_3\approx 0.6$, and then flattens out towards higher values of
$r_2$ or $r_3$. When $r_3\ll r_2$ it oscillates due to the Whittaker function:
for the diagram (i) it peaks at $r_2=r_3\approx 0.6$ and goes to zero in the squeezed limit 
with a damped oscillation. For the diagram (ii), which is sub-dominant
(but is within an order of magnitude of the diagram (i)), it peaks at the equilateral limit and approaches zero in the squeezed limit, also with a damped oscillation.

Similarity of two shapes of the bispectrum can be quantified using a
cosine  defined as $B_h\cdot F^{\rm eq}/\sqrt{(B_h\cdot B_h)(F^{\rm
eq}\cdot F^{\rm eq})}$ \cite{babich/creminelli/zaldarriaga:2004}, where
dot-products denote $X\cdot Y\equiv
\int_0^1dr_2\int_{1-r_2}^1dr_3(r_2r_3)^4X(1,r_2,r_3)Y(1,r_2,r_3)$. We
find 0.89 for the above model parameters, which implies that, despite
the differences in details, these two shapes are similar enough, and
using the equilateral template would be sufficient for the data
analysis, at least for the first trial.

As the sourced $B_h$ and $P_h^2$ have similar exponential dependence on
$m_Q$, we take the ratio to reduce the $m_Q$ dependence. At the
equilateral configuration $k_1=k_2=k_3$ we find
\begin{equation}
\frac{B_h^{RRR}}{P_h^2} \approx \frac{50}{\epsilon_B}\simeq
 \frac{25}{\Omega_A}\,,
\end{equation}
for $3\lesssim m_Q \lesssim 4$. The exact numerical factor multiplying
$\epsilon_B^{-1}$ depends weakly on $m_Q$. This is much greater than
that of the vacuum contribution, $B_h^{\rm vac}/(P_h^{\rm vac})^2$ of order unity \cite{maldacena:2002,maldacena/pimentel:2011}.

This result, i.e., $B_h/P_h^2\propto \Omega_A^{-1}$, may seem strange at
first, as it does not vanish in the absence of the SU(2) field. However,
this result applies only when the sourced GW power spectrum dominates
over the vacuum contribution. This condition does not hold when
$\Omega_A$ is too small, in which case the above result does not apply.
This resembles the situation for the scalar bispectrum in the curvaton scenario
\cite{lyth/wands:2003}.

\section{V. Conclusions}
In summary, we have calculated the bispectrum of GWs sourced by an SU(2)
gauge field coupled to a spectator axion field, finding a large
value.The SU(2) field can also contribute to $\zeta$. In flat gauge $\zeta$ is
given by 
\begin{equation}
\zeta=\frac{\sum_i\delta\rho_i}{3\sum_i(\rho_i+P_i)}
\approx\frac{\Omega_\phi\delta\rho_\phi/\rho_\phi+\Omega_\chi\delta\rho_\chi/\rho_\chi+\Omega_A\delta\rho_A/\rho_A}{2\epsilon}\,,
\end{equation}
where $\rho_i$, $P_i$, $\delta\rho_i$, and $\Omega_i=\rho_i/3H^2M_{\rm Pl}^2$ are the energy density, pressure, energy density perturbation, and energy density
fraction of $i=(\phi,\chi,A)$. We have used
$\rho_\phi+P_\phi=\dot\phi^2=2\epsilon H^2M_{\rm Pl}^2$ and 
ignored the terms related to the axion and gauge fields in the
denominator.  The third term in the numerator is suppressed by
$\Omega_A$ and will be negligible once $\chi$ settles into the potential
minimum and stops producing SU(2), i.e., $\Omega_A\to 0$. The SU(2)
field produces axion perturbations via $t_{ij}+t_{ij}\to \delta\chi$
which, in turn, produces $\zeta$ in two ways. One is via the second term
in the numerator, and is negligible after inflation as $\Omega_\chi\to
0$. Another channel is production of $\delta\phi$ from $\delta\chi$,
producing $\zeta$ via the first term in the numerator. This can in
principle make a sizeable contribution if $m_Q$ is large; however, for
our choice of $3<m_Q<4$ the contribution is several orders of magnitude
smaller than the vacuum contribution of inflaton. We give a rough
order estimate below, and present details in a forthcoming paper
\cite{fujita/namba:prep}.

Considering the exponential dependence $t_{ij}\approx e^{2m_Q}$ (which
includes $m_Q$ dependence of the Whittaker function) and a
vertex $g\Lambda/2(\partial_\eta \delta\chi) t_{ij}t_{ij}$ (where $\Lambda \equiv \lambda Q/f$~\cite{obata/soda:2016,dimastrogiovanni/fasiello/fujita:2016}), the power spectrum of
$\chi$ is evaluated as $k^3P_{\delta\chi}^{tt}/H^2\approx g^2 (\Lambda/2)^2e^{8m_Q}$.
(The vertices that are not proportional to $g$ do not involve
one $\delta\chi$ and two tensors.)
In addition, the gravitational coupling between $\delta\phi$ and $\delta\chi$
is suppressed by $\sqrt{\epsilon_\phi\epsilon_\chi}$. Then we obtain
$k^3P_{\delta\phi}^{tt}/H^{2}\approx \epsilon_\phi \epsilon_\chi g^2 (\Lambda/2)^2
e^{8m_Q}\approx 7.5\times10^{-3}$, where $\epsilon_\phi=10^{-4}$,
$\epsilon_\chi=10^{-8}$, $g=10^{-2}$, and $m_Q=3.45$
(\cite{dimastrogiovanni/fasiello/fujita:2016}, see eq.(3.2) and
Fig.2). Therefore this model can produce a significant amount of GWs
over a wide range in wavenumbers, while simultaneously satisfying
stringent observational constraints on the scalar curvature power
spectrum. 

So far, there is no evidence for primordial non-Gaussianity in both
scalar and tensor perturbations. The Planck collaboration reports limits
on the tensor bispectrum \cite{planckNG:2015} (also see
\cite{shiraishi/liguori/fergusson:2015} for the WMAP limit) in terms of
the following quantity at the equilateral configuration: $f_{\rm
NL}^{\rm tens}\equiv {B^{+++}_{h}(k,k,k)}/{F_\zeta^{\rm eq}(k,k,k)}$, where
$F_\zeta^{\rm eq}(k,k,k)= (18/5)P_\zeta^2(k)$ with $P_\zeta$ being the power
spectrum of the scalar curvature perturbation. Here,
$B_h^{+++}$ is the bispectrum of tensor modes whose polarization tensor
is normalized such that $e_{ij}^+(\bm{k}) e_{ij}^+(-\bm{k})=2$; thus,
$B_h^{+++}$ is related to our bispectrum convention as
$B_h^{+++}=B_h^{RRR}/2\sqrt{2}$. The model then predicts
\begin{equation}
 f_{\rm NL}^{\rm tens}\approx \frac{125}{18\sqrt{2}}\frac{r^2}{\epsilon_B}\approx
  2.5\frac{r^2}{\Omega_A}\,,
\end{equation}
for $3\lesssim m_Q\lesssim 4$, and $r=P_h/P_\zeta$ where $P_\zeta$ is
dominated by the vacuum contribution. As $P_h\propto \epsilon_B$
(Eq.~\eqref{eq:Ph}), $f_{\rm NL}^{\rm tens}$ is proportional to
$\epsilon_B$, hence $\Omega_A$.

As only right-handed modes are amplified, not only the usual
parity-even CMB bispectrum but also a parity-odd bispectrum is produced
\cite{kamionkowski/souradeep:2011,shiraishi/ricciardone/saga:2013}. While
a parity-odd CMB bispectrum gives a clean signature of GWs sourced
by gauge fields, the observational limits are stronger for the
parity-even bispectrum. We thus use the limits combining both bispectra. The
Planck collaboration obtains $f_{\rm NL}^{\rm tens}=400\pm 1500$
(68\%~CL). It is clear that the model predicts
a large tensor bispectrum that is observationally relevant. The planned
future experiments that measure temperature and polarization of the CMB
over full sky, such as LiteBIRD \cite{matsumura/etal:2013}, will tighten
the constraint on, or discover non-zero value of, $f_{\rm NL}^{\rm
tens}$, offering an important test of the origin of primordial GWs
\cite{shiraishi/etal:2016}: is it from vacuum fluctuations, or from sources?

\section{ACKNOWLEDGMENT}
\begin{acknowledgments}
We would like to thank Ryo Namba and Maresuke Shiraishi for useful
 discussions. TF acknowledges the support by Grant-in-Aid for JSPS
 Fellows No.~29-9103. This work was supported also in part by JSPS
 KAKENHI Grant Number JP15H05896.
\end{acknowledgments}
\bibliography{references}

\begin{thebibliography}{48}%
\makeatletter
\providecommand \@ifxundefined [1]{%
 \@ifx{#1\undefined}
}%
\providecommand \@ifnum [1]{%
 \ifnum #1\expandafter \@firstoftwo
 \else \expandafter \@secondoftwo
 \fi
}%
\providecommand \@ifx [1]{%
 \ifx #1\expandafter \@firstoftwo
 \else \expandafter \@secondoftwo
 \fi
}%
\providecommand \natexlab [1]{#1}%
\providecommand \enquote  [1]{``#1''}%
\providecommand \bibnamefont  [1]{#1}%
\providecommand \bibfnamefont [1]{#1}%
\providecommand \citenamefont [1]{#1}%
\providecommand \href@noop [0]{\@secondoftwo}%
\providecommand \href [0]{\begingroup \@sanitize@url \@href}%
\providecommand \@href[1]{\@@startlink{#1}\@@href}%
\providecommand \@@href[1]{\endgroup#1\@@endlink}%
\providecommand \@sanitize@url [0]{\catcode `\\12\catcode `\$12\catcode
  `\&12\catcode `\#12\catcode `\^12\catcode `\_12\catcode `\%12\relax}%
\providecommand \@@startlink[1]{}%
\providecommand \@@endlink[0]{}%
\providecommand \url  [0]{\begingroup\@sanitize@url \@url }%
\providecommand \@url [1]{\endgroup\@href {#1}{\urlprefix }}%
\providecommand \urlprefix  [0]{URL }%
\providecommand \Eprint [0]{\href }%
\providecommand \doibase [0]{http://dx.doi.org/}%
\providecommand \selectlanguage [0]{\@gobble}%
\providecommand \bibinfo  [0]{\@secondoftwo}%
\providecommand \bibfield  [0]{\@secondoftwo}%
\providecommand \translation [1]{[#1]}%
\providecommand \BibitemOpen [0]{}%
\providecommand \bibitemStop [0]{}%
\providecommand \bibitemNoStop [0]{.\EOS\space}%
\providecommand \EOS [0]{\spacefactor3000\relax}%
\providecommand \BibitemShut  [1]{\csname bibitem#1\endcsname}%
\let\auto@bib@innerbib\@empty
\bibitem [{\citenamefont {Sato}(1981)}]{sato:1981}%
  \BibitemOpen
  \bibfield  {author} {\bibinfo {author} {\bibfnamefont {K.}~\bibnamefont
  {Sato}},\ }\href@noop {} {\bibfield  {journal} {\bibinfo  {journal} {Mon.
  Not. Roy. Astron. Soc.}\ }\textbf {\bibinfo {volume} {195}},\ \bibinfo
  {pages} {467} (\bibinfo {year} {1981})}\BibitemShut {NoStop}%
\bibitem [{\citenamefont {Guth}(1981)}]{guth:1981}%
  \BibitemOpen
  \bibfield  {author} {\bibinfo {author} {\bibfnamefont {A.~H.}\ \bibnamefont
  {Guth}},\ }\href {\doibase 10.1103/PhysRevD.23.347} {\bibfield  {journal}
  {\bibinfo  {journal} {Phys. Rev.}\ }\textbf {\bibinfo {volume} {D23}},\
  \bibinfo {pages} {347} (\bibinfo {year} {1981})}\BibitemShut {NoStop}%
\bibitem [{\citenamefont {Linde}(1982)}]{linde:1982}%
  \BibitemOpen
  \bibfield  {author} {\bibinfo {author} {\bibfnamefont {A.~D.}\ \bibnamefont
  {Linde}},\ }\href {\doibase 10.1016/0370-2693(82)91219-9} {\bibfield
  {journal} {\bibinfo  {journal} {Phys. Lett.}\ }\textbf {\bibinfo {volume}
  {B108}},\ \bibinfo {pages} {389} (\bibinfo {year} {1982})}\BibitemShut
  {NoStop}%
\bibitem [{\citenamefont {Albrecht}\ and\ \citenamefont
  {Steinhardt}(1982)}]{albrecht/steinhardt:1982}%
  \BibitemOpen
  \bibfield  {author} {\bibinfo {author} {\bibfnamefont {A.}~\bibnamefont
  {Albrecht}}\ and\ \bibinfo {author} {\bibfnamefont {P.~J.}\ \bibnamefont
  {Steinhardt}},\ }\href {\doibase 10.1103/PhysRevLett.48.1220} {\bibfield
  {journal} {\bibinfo  {journal} {Phys. Rev. Lett.}\ }\textbf {\bibinfo
  {volume} {48}},\ \bibinfo {pages} {1220} (\bibinfo {year}
  {1982})}\BibitemShut {NoStop}%
\bibitem [{\citenamefont {Grishchuk}(1975)}]{grishchuk:1974}%
  \BibitemOpen
  \bibfield  {author} {\bibinfo {author} {\bibfnamefont {L.~P.}\ \bibnamefont
  {Grishchuk}},\ }\href@noop {} {\bibfield  {journal} {\bibinfo  {journal}
  {Sov. Phys. JETP}\ }\textbf {\bibinfo {volume} {40}},\ \bibinfo {pages} {409}
  (\bibinfo {year} {1975})},\ \bibinfo {note} {[Zh. Eksp. Teor.
  Fiz.67,825(1974)]}\BibitemShut {NoStop}%
\bibitem [{\citenamefont {Starobinsky}(1979)}]{starobinsky:1979}%
  \BibitemOpen
  \bibfield  {author} {\bibinfo {author} {\bibfnamefont {A.~A.}\ \bibnamefont
  {Starobinsky}},\ }\href@noop {} {\bibfield  {journal} {\bibinfo  {journal}
  {JETP Lett.}\ }\textbf {\bibinfo {volume} {30}},\ \bibinfo {pages} {682}
  (\bibinfo {year} {1979})},\ \bibinfo {note} {[Pisma Zh. Eksp. Teor.
  Fiz.30,719(1979)]}\BibitemShut {NoStop}%
\bibitem [{\citenamefont {Abbott}\ and\ \citenamefont
  {Wise}(1984)}]{abbott/wise:1984}%
  \BibitemOpen
  \bibfield  {author} {\bibinfo {author} {\bibfnamefont {L.~F.}\ \bibnamefont
  {Abbott}}\ and\ \bibinfo {author} {\bibfnamefont {M.~B.}\ \bibnamefont
  {Wise}},\ }\href {\doibase 10.1016/0550-3213(84)90329-8} {\bibfield
  {journal} {\bibinfo  {journal} {Nucl. Phys.}\ }\textbf {\bibinfo {volume}
  {B244}},\ \bibinfo {pages} {541} (\bibinfo {year} {1984})}\BibitemShut
  {NoStop}%
\bibitem [{\citenamefont {Rubakov}\ \emph {et~al.}(1982)\citenamefont
  {Rubakov}, \citenamefont {Sazhin},\ and\ \citenamefont
  {Veryaskin}}]{rubakov/sazhin/veryaskin:1982}%
  \BibitemOpen
  \bibfield  {author} {\bibinfo {author} {\bibfnamefont {V.~A.}\ \bibnamefont
  {Rubakov}}, \bibinfo {author} {\bibfnamefont {M.~V.}\ \bibnamefont {Sazhin}},
  \ and\ \bibinfo {author} {\bibfnamefont {A.~V.}\ \bibnamefont {Veryaskin}},\
  }\href {\doibase 10.1016/0370-2693(82)90641-4} {\bibfield  {journal}
  {\bibinfo  {journal} {Phys. Lett.}\ }\textbf {\bibinfo {volume} {B115}},\
  \bibinfo {pages} {189} (\bibinfo {year} {1982})}\BibitemShut {NoStop}%
\bibitem [{\citenamefont {Fabbri}\ and\ \citenamefont
  {Pollock}(1983)}]{fabbri/pollock:1983}%
  \BibitemOpen
  \bibfield  {author} {\bibinfo {author} {\bibfnamefont {R.}~\bibnamefont
  {Fabbri}}\ and\ \bibinfo {author} {\bibfnamefont {M.~d.}\ \bibnamefont
  {Pollock}},\ }\href {\doibase 10.1016/0370-2693(83)91322-9} {\bibfield
  {journal} {\bibinfo  {journal} {Phys. Lett.}\ }\textbf {\bibinfo {volume}
  {B125}},\ \bibinfo {pages} {445} (\bibinfo {year} {1983})}\BibitemShut
  {NoStop}%
\bibitem [{\citenamefont {Starobinsky}(1985)}]{starobinsky:1985}%
  \BibitemOpen
  \bibfield  {author} {\bibinfo {author} {\bibfnamefont {A.~A.}\ \bibnamefont
  {Starobinsky}},\ }\href@noop {} {\bibfield  {journal} {\bibinfo  {journal}
  {Sov. Astron. Lett.}\ }\textbf {\bibinfo {volume} {11}},\ \bibinfo {pages}
  {133} (\bibinfo {year} {1985})}\BibitemShut {NoStop}%
\bibitem [{\citenamefont {{Polnarev}}(1985)}]{polnarev:1985}%
  \BibitemOpen
  \bibfield  {author} {\bibinfo {author} {\bibfnamefont {A.~G.}\ \bibnamefont
  {{Polnarev}}},\ }\href@noop {} {\bibfield  {journal} {\bibinfo  {journal}
  {Sov. Astron.}\ }\textbf {\bibinfo {volume} {29}},\ \bibinfo {pages} {607}
  (\bibinfo {year} {1985})}\BibitemShut {NoStop}%
\bibitem [{\citenamefont {Crittenden}\ \emph {et~al.}(1993)\citenamefont
  {Crittenden}, \citenamefont {Davis},\ and\ \citenamefont
  {Steinhardt}}]{crittenden/davis/steinhardt:1993}%
  \BibitemOpen
  \bibfield  {author} {\bibinfo {author} {\bibfnamefont {R.}~\bibnamefont
  {Crittenden}}, \bibinfo {author} {\bibfnamefont {R.~L.}\ \bibnamefont
  {Davis}}, \ and\ \bibinfo {author} {\bibfnamefont {P.~J.}\ \bibnamefont
  {Steinhardt}},\ }\href {\doibase 10.1086/187082} {\bibfield  {journal}
  {\bibinfo  {journal} {Astrophys. J.}\ }\textbf {\bibinfo {volume} {417}},\
  \bibinfo {pages} {L13} (\bibinfo {year} {1993})},\ \Eprint
  {http://arxiv.org/abs/astro-ph/9306027} {arXiv:astro-ph/9306027 [astro-ph]}
  \BibitemShut {NoStop}%
\bibitem [{\citenamefont {Seljak}\ and\ \citenamefont
  {Zaldarriaga}(1997)}]{seljak/zaldarriaga:1996}%
  \BibitemOpen
  \bibfield  {author} {\bibinfo {author} {\bibfnamefont {U.}~\bibnamefont
  {Seljak}}\ and\ \bibinfo {author} {\bibfnamefont {M.}~\bibnamefont
  {Zaldarriaga}},\ }\href {\doibase 10.1103/PhysRevLett.78.2054} {\bibfield
  {journal} {\bibinfo  {journal} {Phys. Rev. Lett.}\ }\textbf {\bibinfo
  {volume} {78}},\ \bibinfo {pages} {2054} (\bibinfo {year} {1997})},\ \Eprint
  {http://arxiv.org/abs/astro-ph/9609169} {arXiv:astro-ph/9609169 [astro-ph]}
  \BibitemShut {NoStop}%
\bibitem [{\citenamefont {Kamionkowski}\ \emph {et~al.}(1997)\citenamefont
  {Kamionkowski}, \citenamefont {Kosowsky},\ and\ \citenamefont
  {Stebbins}}]{kamionkowski/kosowsky/stebbins:1996}%
  \BibitemOpen
  \bibfield  {author} {\bibinfo {author} {\bibfnamefont {M.}~\bibnamefont
  {Kamionkowski}}, \bibinfo {author} {\bibfnamefont {A.}~\bibnamefont
  {Kosowsky}}, \ and\ \bibinfo {author} {\bibfnamefont {A.}~\bibnamefont
  {Stebbins}},\ }\href {\doibase 10.1103/PhysRevLett.78.2058} {\bibfield
  {journal} {\bibinfo  {journal} {Phys. Rev. Lett.}\ }\textbf {\bibinfo
  {volume} {78}},\ \bibinfo {pages} {2058} (\bibinfo {year} {1997})},\ \Eprint
  {http://arxiv.org/abs/astro-ph/9609132} {arXiv:astro-ph/9609132 [astro-ph]}
  \BibitemShut {NoStop}%
\bibitem [{\citenamefont {Cook}\ and\ \citenamefont
  {Sorbo}(2012)}]{cook/sorbo:2012}%
  \BibitemOpen
  \bibfield  {author} {\bibinfo {author} {\bibfnamefont {J.~L.}\ \bibnamefont
  {Cook}}\ and\ \bibinfo {author} {\bibfnamefont {L.}~\bibnamefont {Sorbo}},\
  }\href {\doibase 10.1103/PhysRevD.86.069901, 10.1103/PhysRevD.85.023534}
  {\bibfield  {journal} {\bibinfo  {journal} {Phys. Rev.}\ }\textbf {\bibinfo
  {volume} {D85}},\ \bibinfo {pages} {023534} (\bibinfo {year} {2012})},\
  \bibinfo {note} {[Erratum: Phys. Rev.D86,069901(2012)]},\ \Eprint
  {http://arxiv.org/abs/1109.0022} {arXiv:1109.0022 [astro-ph.CO]} \BibitemShut
  {NoStop}%
\bibitem [{\citenamefont {Carney}\ \emph {et~al.}(2012)\citenamefont {Carney},
  \citenamefont {Fischler}, \citenamefont {Kovetz}, \citenamefont
  {Lorshbough},\ and\ \citenamefont {Paban}}]{carney/etal:2012}%
  \BibitemOpen
  \bibfield  {author} {\bibinfo {author} {\bibfnamefont {D.}~\bibnamefont
  {Carney}}, \bibinfo {author} {\bibfnamefont {W.}~\bibnamefont {Fischler}},
  \bibinfo {author} {\bibfnamefont {E.~D.}\ \bibnamefont {Kovetz}}, \bibinfo
  {author} {\bibfnamefont {D.}~\bibnamefont {Lorshbough}}, \ and\ \bibinfo
  {author} {\bibfnamefont {S.}~\bibnamefont {Paban}},\ }\href {\doibase
  10.1007/JHEP11(2012)042} {\bibfield  {journal} {\bibinfo  {journal} {JHEP}\
  }\textbf {\bibinfo {volume} {11}},\ \bibinfo {pages} {042} (\bibinfo {year}
  {2012})},\ \Eprint {http://arxiv.org/abs/1209.3848} {arXiv:1209.3848
  [hep-th]} \BibitemShut {NoStop}%
\bibitem [{\citenamefont {Biagetti}\ \emph {et~al.}(2013)\citenamefont
  {Biagetti}, \citenamefont {Fasiello},\ and\ \citenamefont
  {Riotto}}]{biagetti/fasiello/riotto:2013}%
  \BibitemOpen
  \bibfield  {author} {\bibinfo {author} {\bibfnamefont {M.}~\bibnamefont
  {Biagetti}}, \bibinfo {author} {\bibfnamefont {M.}~\bibnamefont {Fasiello}},
  \ and\ \bibinfo {author} {\bibfnamefont {A.}~\bibnamefont {Riotto}},\ }\href
  {\doibase 10.1103/PhysRevD.88.103518} {\bibfield  {journal} {\bibinfo
  {journal} {Phys. Rev.}\ }\textbf {\bibinfo {volume} {D88}},\ \bibinfo {pages}
  {103518} (\bibinfo {year} {2013})},\ \Eprint {http://arxiv.org/abs/1305.7241}
  {arXiv:1305.7241 [astro-ph.CO]} \BibitemShut {NoStop}%
\bibitem [{\citenamefont {Senatore}\ \emph {et~al.}(2014)\citenamefont
  {Senatore}, \citenamefont {Silverstein},\ and\ \citenamefont
  {Zaldarriaga}}]{senatore/silverstein/zaldarriaga:2014}%
  \BibitemOpen
  \bibfield  {author} {\bibinfo {author} {\bibfnamefont {L.}~\bibnamefont
  {Senatore}}, \bibinfo {author} {\bibfnamefont {E.}~\bibnamefont
  {Silverstein}}, \ and\ \bibinfo {author} {\bibfnamefont {M.}~\bibnamefont
  {Zaldarriaga}},\ }\href {\doibase 10.1088/1475-7516/2014/08/016} {\bibfield
  {journal} {\bibinfo  {journal} {JCAP}\ }\textbf {\bibinfo {volume} {1408}},\
  \bibinfo {pages} {016} (\bibinfo {year} {2014})},\ \Eprint
  {http://arxiv.org/abs/1109.0542} {arXiv:1109.0542 [hep-th]} \BibitemShut
  {NoStop}%
\bibitem [{\citenamefont {Sorbo}(2011)}]{sorbo:2011}%
  \BibitemOpen
  \bibfield  {author} {\bibinfo {author} {\bibfnamefont {L.}~\bibnamefont
  {Sorbo}},\ }\href {\doibase 10.1088/1475-7516/2011/06/003} {\bibfield
  {journal} {\bibinfo  {journal} {JCAP}\ }\textbf {\bibinfo {volume} {1106}},\
  \bibinfo {pages} {003} (\bibinfo {year} {2011})},\ \Eprint
  {http://arxiv.org/abs/1101.1525} {arXiv:1101.1525 [astro-ph.CO]} \BibitemShut
  {NoStop}%
\bibitem [{\citenamefont {Anber}\ and\ \citenamefont
  {Sorbo}(2012)}]{anber/sorbo:2012}%
  \BibitemOpen
  \bibfield  {author} {\bibinfo {author} {\bibfnamefont {M.~M.}\ \bibnamefont
  {Anber}}\ and\ \bibinfo {author} {\bibfnamefont {L.}~\bibnamefont {Sorbo}},\
  }\href {\doibase 10.1103/PhysRevD.85.123537} {\bibfield  {journal} {\bibinfo
  {journal} {Phys. Rev.}\ }\textbf {\bibinfo {volume} {D85}},\ \bibinfo {pages}
  {123537} (\bibinfo {year} {2012})},\ \Eprint {http://arxiv.org/abs/1203.5849}
  {arXiv:1203.5849 [astro-ph.CO]} \BibitemShut {NoStop}%
\bibitem [{\citenamefont {Barnaby}\ and\ \citenamefont
  {Peloso}(2011)}]{barnaby/peloso:2011}%
  \BibitemOpen
  \bibfield  {author} {\bibinfo {author} {\bibfnamefont {N.}~\bibnamefont
  {Barnaby}}\ and\ \bibinfo {author} {\bibfnamefont {M.}~\bibnamefont
  {Peloso}},\ }\href {\doibase 10.1103/PhysRevLett.106.181301} {\bibfield
  {journal} {\bibinfo  {journal} {Phys. Rev. Lett.}\ }\textbf {\bibinfo
  {volume} {106}},\ \bibinfo {pages} {181301} (\bibinfo {year} {2011})},\
  \Eprint {http://arxiv.org/abs/1011.1500} {arXiv:1011.1500 [hep-ph]}
  \BibitemShut {NoStop}%
\bibitem [{\citenamefont {Barnaby}\ \emph {et~al.}(2012)\citenamefont
  {Barnaby}, \citenamefont {Moxon}, \citenamefont {Namba}, \citenamefont
  {Peloso}, \citenamefont {Shiu},\ and\ \citenamefont
  {Zhou}}]{barnaby/etal:2012}%
  \BibitemOpen
  \bibfield  {author} {\bibinfo {author} {\bibfnamefont {N.}~\bibnamefont
  {Barnaby}}, \bibinfo {author} {\bibfnamefont {J.}~\bibnamefont {Moxon}},
  \bibinfo {author} {\bibfnamefont {R.}~\bibnamefont {Namba}}, \bibinfo
  {author} {\bibfnamefont {M.}~\bibnamefont {Peloso}}, \bibinfo {author}
  {\bibfnamefont {G.}~\bibnamefont {Shiu}}, \ and\ \bibinfo {author}
  {\bibfnamefont {P.}~\bibnamefont {Zhou}},\ }\href {\doibase
  10.1103/PhysRevD.86.103508} {\bibfield  {journal} {\bibinfo  {journal} {Phys.
  Rev.}\ }\textbf {\bibinfo {volume} {D86}},\ \bibinfo {pages} {103508}
  (\bibinfo {year} {2012})},\ \Eprint {http://arxiv.org/abs/1206.6117}
  {arXiv:1206.6117 [astro-ph.CO]} \BibitemShut {NoStop}%
\bibitem [{\citenamefont {Peloso}\ \emph {et~al.}(2016)\citenamefont {Peloso},
  \citenamefont {Sorbo},\ and\ \citenamefont {Unal}}]{peloso/sorbo/unal:2016}%
  \BibitemOpen
  \bibfield  {author} {\bibinfo {author} {\bibfnamefont {M.}~\bibnamefont
  {Peloso}}, \bibinfo {author} {\bibfnamefont {L.}~\bibnamefont {Sorbo}}, \
  and\ \bibinfo {author} {\bibfnamefont {C.}~\bibnamefont {Unal}},\ }\href
  {\doibase 10.1088/1475-7516/2016/09/001} {\bibfield  {journal} {\bibinfo
  {journal} {JCAP}\ }\textbf {\bibinfo {volume} {1609}},\ \bibinfo {pages}
  {001} (\bibinfo {year} {2016})},\ \Eprint {http://arxiv.org/abs/1606.00459}
  {arXiv:1606.00459 [astro-ph.CO]} \BibitemShut {NoStop}%
\bibitem [{\citenamefont {Maleknejad}\ and\ \citenamefont
  {Sheikh-Jabbari}(2013)}]{maleknejad/sheikh-jabbari:2013}%
  \BibitemOpen
  \bibfield  {author} {\bibinfo {author} {\bibfnamefont {A.}~\bibnamefont
  {Maleknejad}}\ and\ \bibinfo {author} {\bibfnamefont {M.~M.}\ \bibnamefont
  {Sheikh-Jabbari}},\ }\href {\doibase 10.1016/j.physletb.2013.05.001}
  {\bibfield  {journal} {\bibinfo  {journal} {Phys. Lett.}\ }\textbf {\bibinfo
  {volume} {B723}},\ \bibinfo {pages} {224} (\bibinfo {year} {2013})},\ \Eprint
  {http://arxiv.org/abs/1102.1513} {arXiv:1102.1513 [hep-ph]} \BibitemShut
  {NoStop}%
\bibitem [{\citenamefont {Dimastrogiovanni}\ and\ \citenamefont
  {Peloso}(2013)}]{dimastrogiovanni/peloso:2012}%
  \BibitemOpen
  \bibfield  {author} {\bibinfo {author} {\bibfnamefont {E.}~\bibnamefont
  {Dimastrogiovanni}}\ and\ \bibinfo {author} {\bibfnamefont {M.}~\bibnamefont
  {Peloso}},\ }\href {\doibase 10.1103/PhysRevD.87.103501} {\bibfield
  {journal} {\bibinfo  {journal} {Phys. Rev.}\ }\textbf {\bibinfo {volume}
  {D87}},\ \bibinfo {pages} {103501} (\bibinfo {year} {2013})},\ \Eprint
  {http://arxiv.org/abs/1212.5184} {arXiv:1212.5184 [astro-ph.CO]} \BibitemShut
  {NoStop}%
\bibitem [{\citenamefont {Adshead}\ \emph
  {et~al.}(2013{\natexlab{a}})\citenamefont {Adshead}, \citenamefont
  {Martinec},\ and\ \citenamefont {Wyman}}]{adshead/etal:2013}%
  \BibitemOpen
  \bibfield  {author} {\bibinfo {author} {\bibfnamefont {P.}~\bibnamefont
  {Adshead}}, \bibinfo {author} {\bibfnamefont {E.}~\bibnamefont {Martinec}}, \
  and\ \bibinfo {author} {\bibfnamefont {M.}~\bibnamefont {Wyman}},\ }\href
  {\doibase 10.1007/JHEP09(2013)087} {\bibfield  {journal} {\bibinfo  {journal}
  {JHEP}\ }\textbf {\bibinfo {volume} {09}},\ \bibinfo {pages} {087} (\bibinfo
  {year} {2013}{\natexlab{a}})},\ \Eprint {http://arxiv.org/abs/1305.2930}
  {arXiv:1305.2930 [hep-th]} \BibitemShut {NoStop}%
\bibitem [{\citenamefont {Adshead}\ \emph
  {et~al.}(2013{\natexlab{b}})\citenamefont {Adshead}, \citenamefont
  {Martinec},\ and\ \citenamefont {Wyman}}]{adshead/martinec/wyman:2013}%
  \BibitemOpen
  \bibfield  {author} {\bibinfo {author} {\bibfnamefont {P.}~\bibnamefont
  {Adshead}}, \bibinfo {author} {\bibfnamefont {E.}~\bibnamefont {Martinec}}, \
  and\ \bibinfo {author} {\bibfnamefont {M.}~\bibnamefont {Wyman}},\ }\href
  {\doibase 10.1103/PhysRevD.88.021302} {\bibfield  {journal} {\bibinfo
  {journal} {Phys. Rev.}\ }\textbf {\bibinfo {volume} {D88}},\ \bibinfo {pages}
  {021302} (\bibinfo {year} {2013}{\natexlab{b}})},\ \Eprint
  {http://arxiv.org/abs/1301.2598} {arXiv:1301.2598 [hep-th]} \BibitemShut
  {NoStop}%
\bibitem [{\citenamefont {Maleknejad}(2016)}]{maleknejad:2016}%
  \BibitemOpen
  \bibfield  {author} {\bibinfo {author} {\bibfnamefont {A.}~\bibnamefont
  {Maleknejad}},\ }\href {\doibase 10.1007/JHEP07(2016)104} {\bibfield
  {journal} {\bibinfo  {journal} {JHEP}\ }\textbf {\bibinfo {volume} {07}},\
  \bibinfo {pages} {104} (\bibinfo {year} {2016})},\ \Eprint
  {http://arxiv.org/abs/1604.03327} {arXiv:1604.03327 [hep-ph]} \BibitemShut
  {NoStop}%
\bibitem [{\citenamefont {Dimastrogiovanni}\ \emph {et~al.}(2017)\citenamefont
  {Dimastrogiovanni}, \citenamefont {Fasiello},\ and\ \citenamefont
  {Fujita}}]{dimastrogiovanni/fasiello/fujita:2016}%
  \BibitemOpen
  \bibfield  {author} {\bibinfo {author} {\bibfnamefont {E.}~\bibnamefont
  {Dimastrogiovanni}}, \bibinfo {author} {\bibfnamefont {M.}~\bibnamefont
  {Fasiello}}, \ and\ \bibinfo {author} {\bibfnamefont {T.}~\bibnamefont
  {Fujita}},\ }\href {\doibase 10.1088/1475-7516/2017/01/019} {\bibfield
  {journal} {\bibinfo  {journal} {JCAP}\ }\textbf {\bibinfo {volume} {1701}},\
  \bibinfo {pages} {019} (\bibinfo {year} {2017})},\ \Eprint
  {http://arxiv.org/abs/1608.04216} {arXiv:1608.04216 [astro-ph.CO]}
  \BibitemShut {NoStop}%
\bibitem [{\citenamefont {Cook}\ and\ \citenamefont
  {Sorbo}(2013)}]{cook/sorbo:2013}%
  \BibitemOpen
  \bibfield  {author} {\bibinfo {author} {\bibfnamefont {J.~L.}\ \bibnamefont
  {Cook}}\ and\ \bibinfo {author} {\bibfnamefont {L.}~\bibnamefont {Sorbo}},\
  }\href {\doibase 10.1088/1475-7516/2013/11/047} {\bibfield  {journal}
  {\bibinfo  {journal} {JCAP}\ }\textbf {\bibinfo {volume} {1311}},\ \bibinfo
  {pages} {047} (\bibinfo {year} {2013})},\ \Eprint
  {http://arxiv.org/abs/1307.7077} {arXiv:1307.7077 [astro-ph.CO]} \BibitemShut
  {NoStop}%
\bibitem [{\citenamefont {Ferreira}\ and\ \citenamefont
  {Sloth}(2014)}]{ferreira/sloth:2014}%
  \BibitemOpen
  \bibfield  {author} {\bibinfo {author} {\bibfnamefont {R.~Z.}\ \bibnamefont
  {Ferreira}}\ and\ \bibinfo {author} {\bibfnamefont {M.~S.}\ \bibnamefont
  {Sloth}},\ }\href {\doibase 10.1007/JHEP12(2014)139} {\bibfield  {journal}
  {\bibinfo  {journal} {JHEP}\ }\textbf {\bibinfo {volume} {12}},\ \bibinfo
  {pages} {139} (\bibinfo {year} {2014})},\ \Eprint
  {http://arxiv.org/abs/1409.5799} {arXiv:1409.5799 [hep-ph]} \BibitemShut
  {NoStop}%
\bibitem [{\citenamefont {Namba}\ \emph {et~al.}(2016)\citenamefont {Namba},
  \citenamefont {Peloso}, \citenamefont {Shiraishi}, \citenamefont {Sorbo},\
  and\ \citenamefont {Unal}}]{namba/etal:2015}%
  \BibitemOpen
  \bibfield  {author} {\bibinfo {author} {\bibfnamefont {R.}~\bibnamefont
  {Namba}}, \bibinfo {author} {\bibfnamefont {M.}~\bibnamefont {Peloso}},
  \bibinfo {author} {\bibfnamefont {M.}~\bibnamefont {Shiraishi}}, \bibinfo
  {author} {\bibfnamefont {L.}~\bibnamefont {Sorbo}}, \ and\ \bibinfo {author}
  {\bibfnamefont {C.}~\bibnamefont {Unal}},\ }\href {\doibase
  10.1088/1475-7516/2016/01/041} {\bibfield  {journal} {\bibinfo  {journal}
  {JCAP}\ }\textbf {\bibinfo {volume} {1601}},\ \bibinfo {pages} {041}
  (\bibinfo {year} {2016})},\ \Eprint {http://arxiv.org/abs/1509.07521}
  {arXiv:1509.07521 [astro-ph.CO]} \BibitemShut {NoStop}%
\bibitem [{\citenamefont {Adshead}\ and\ \citenamefont
  {Wyman}(2012)}]{adshead/wyman:2012}%
  \BibitemOpen
  \bibfield  {author} {\bibinfo {author} {\bibfnamefont {P.}~\bibnamefont
  {Adshead}}\ and\ \bibinfo {author} {\bibfnamefont {M.}~\bibnamefont
  {Wyman}},\ }\href {\doibase 10.1103/PhysRevLett.108.261302} {\bibfield
  {journal} {\bibinfo  {journal} {Phys. Rev. Lett.}\ }\textbf {\bibinfo
  {volume} {108}},\ \bibinfo {pages} {261302} (\bibinfo {year} {2012})},\
  \Eprint {http://arxiv.org/abs/1202.2366} {arXiv:1202.2366 [hep-th]}
  \BibitemShut {NoStop}%
\bibitem [{\citenamefont {Obata}\ and\ \citenamefont
  {Soda}(2016)}]{obata/soda:2016}%
  \BibitemOpen
  \bibfield  {author} {\bibinfo {author} {\bibfnamefont {I.}~\bibnamefont
  {Obata}}\ and\ \bibinfo {author} {\bibfnamefont {J.}~\bibnamefont {Soda}},\
  }\href {\doibase 10.1103/PhysRevD.93.123502} {\bibfield  {journal} {\bibinfo
  {journal} {Phys. Rev.}\ }\textbf {\bibinfo {volume} {D93}},\ \bibinfo {pages}
  {123502} (\bibinfo {year} {2016})},\ \Eprint
  {http://arxiv.org/abs/1602.06024} {arXiv:1602.06024 [hep-th]} \BibitemShut
  {NoStop}%
\bibitem [{\citenamefont {Fujita}\ \emph {et~al.}(2017)\citenamefont {Fujita},
  \citenamefont {Namba},\ and\ \citenamefont {Tada}}]{fujita/namba/tada:prep}%
  \BibitemOpen
  \bibfield  {author} {\bibinfo {author} {\bibfnamefont {T.}~\bibnamefont
  {Fujita}}, \bibinfo {author} {\bibfnamefont {R.}~\bibnamefont {Namba}}, \
  and\ \bibinfo {author} {\bibfnamefont {Y.}~\bibnamefont {Tada}},\ }\href@noop
  {} {\  (\bibinfo {year} {2017})},\ \Eprint {http://arxiv.org/abs/1705.01533}
  {arXiv:1705.01533 [astro-ph.CO]} \BibitemShut {NoStop}%
\bibitem [{\citenamefont {Seery}\ \emph {et~al.}(2008)\citenamefont {Seery},
  \citenamefont {Malik},\ and\ \citenamefont {Lyth}}]{seery:2008}%
  \BibitemOpen
  \bibfield  {author} {\bibinfo {author} {\bibfnamefont {D.}~\bibnamefont
  {Seery}}, \bibinfo {author} {\bibfnamefont {K.~A.}\ \bibnamefont {Malik}}, \
  and\ \bibinfo {author} {\bibfnamefont {D.~H.}\ \bibnamefont {Lyth}},\ }\href
  {\doibase 10.1088/1475-7516/2008/03/014} {\bibfield  {journal} {\bibinfo
  {journal} {JCAP}\ }\textbf {\bibinfo {volume} {0803}},\ \bibinfo {pages}
  {014} (\bibinfo {year} {2008})},\ \Eprint {http://arxiv.org/abs/0802.0588}
  {arXiv:0802.0588 [astro-ph]} \BibitemShut {NoStop}%
\bibitem [{\citenamefont {Creminelli}\ \emph {et~al.}(2006)\citenamefont
  {Creminelli}, \citenamefont {Nicolis}, \citenamefont {Senatore},
  \citenamefont {Tegmark},\ and\ \citenamefont
  {Zaldarriaga}}]{creminelli/etal:2006}%
  \BibitemOpen
  \bibfield  {author} {\bibinfo {author} {\bibfnamefont {P.}~\bibnamefont
  {Creminelli}}, \bibinfo {author} {\bibfnamefont {A.}~\bibnamefont {Nicolis}},
  \bibinfo {author} {\bibfnamefont {L.}~\bibnamefont {Senatore}}, \bibinfo
  {author} {\bibfnamefont {M.}~\bibnamefont {Tegmark}}, \ and\ \bibinfo
  {author} {\bibfnamefont {M.}~\bibnamefont {Zaldarriaga}},\ }\href {\doibase
  10.1088/1475-7516/2006/05/004} {\bibfield  {journal} {\bibinfo  {journal}
  {JCAP}\ }\textbf {\bibinfo {volume} {0605}},\ \bibinfo {pages} {004}
  (\bibinfo {year} {2006})},\ \Eprint {http://arxiv.org/abs/astro-ph/0509029}
  {arXiv:astro-ph/0509029 [astro-ph]} \BibitemShut {NoStop}%
\bibitem [{\citenamefont {Babich}\ \emph {et~al.}(2004)\citenamefont {Babich},
  \citenamefont {Creminelli},\ and\ \citenamefont
  {Zaldarriaga}}]{babich/creminelli/zaldarriaga:2004}%
  \BibitemOpen
  \bibfield  {author} {\bibinfo {author} {\bibfnamefont {D.}~\bibnamefont
  {Babich}}, \bibinfo {author} {\bibfnamefont {P.}~\bibnamefont {Creminelli}},
  \ and\ \bibinfo {author} {\bibfnamefont {M.}~\bibnamefont {Zaldarriaga}},\
  }\href {\doibase 10.1088/1475-7516/2004/08/009} {\bibfield  {journal}
  {\bibinfo  {journal} {JCAP}\ }\textbf {\bibinfo {volume} {0408}},\ \bibinfo
  {pages} {009} (\bibinfo {year} {2004})},\ \Eprint
  {http://arxiv.org/abs/astro-ph/0405356} {arXiv:astro-ph/0405356 [astro-ph]}
  \BibitemShut {NoStop}%
\bibitem [{\citenamefont {Maldacena}(2003)}]{maldacena:2002}%
  \BibitemOpen
  \bibfield  {author} {\bibinfo {author} {\bibfnamefont {J.~M.}\ \bibnamefont
  {Maldacena}},\ }\href {\doibase 10.1088/1126-6708/2003/05/013} {\bibfield
  {journal} {\bibinfo  {journal} {JHEP}\ }\textbf {\bibinfo {volume} {05}},\
  \bibinfo {pages} {013} (\bibinfo {year} {2003})},\ \Eprint
  {http://arxiv.org/abs/astro-ph/0210603} {arXiv:astro-ph/0210603 [astro-ph]}
  \BibitemShut {NoStop}%
\bibitem [{\citenamefont {Maldacena}\ and\ \citenamefont
  {Pimentel}(2011)}]{maldacena/pimentel:2011}%
  \BibitemOpen
  \bibfield  {author} {\bibinfo {author} {\bibfnamefont {J.~M.}\ \bibnamefont
  {Maldacena}}\ and\ \bibinfo {author} {\bibfnamefont {G.~L.}\ \bibnamefont
  {Pimentel}},\ }\href {\doibase 10.1007/JHEP09(2011)045} {\bibfield  {journal}
  {\bibinfo  {journal} {JHEP}\ }\textbf {\bibinfo {volume} {09}},\ \bibinfo
  {pages} {045} (\bibinfo {year} {2011})},\ \Eprint
  {http://arxiv.org/abs/1104.2846} {arXiv:1104.2846 [hep-th]} \BibitemShut
  {NoStop}%
\bibitem [{\citenamefont {Lyth}\ \emph {et~al.}(2003)\citenamefont {Lyth},
  \citenamefont {Ungarelli},\ and\ \citenamefont {Wands}}]{lyth/wands:2003}%
  \BibitemOpen
  \bibfield  {author} {\bibinfo {author} {\bibfnamefont {D.~H.}\ \bibnamefont
  {Lyth}}, \bibinfo {author} {\bibfnamefont {C.}~\bibnamefont {Ungarelli}}, \
  and\ \bibinfo {author} {\bibfnamefont {D.}~\bibnamefont {Wands}},\ }\href
  {\doibase 10.1103/PhysRevD.67.023503} {\bibfield  {journal} {\bibinfo
  {journal} {Phys. Rev.}\ }\textbf {\bibinfo {volume} {D67}},\ \bibinfo {pages}
  {023503} (\bibinfo {year} {2003})},\ \Eprint
  {http://arxiv.org/abs/astro-ph/0208055} {arXiv:astro-ph/0208055 [astro-ph]}
  \BibitemShut {NoStop}%
\bibitem [{\citenamefont {Fujita}\ and\ \citenamefont
  {Namba}()}]{fujita/namba:prep}%
  \BibitemOpen
  \bibfield  {author} {\bibinfo {author} {\bibfnamefont {T.}~\bibnamefont
  {Fujita}}\ and\ \bibinfo {author} {\bibfnamefont {R.}~\bibnamefont {Namba}},\
  }\href@noop {} {\bibinfo  {journal} {in prep.}\ }\BibitemShut {NoStop}%
\bibitem [{\citenamefont {Ade}\ \emph {et~al.}(2016)\citenamefont {Ade} \emph
  {et~al.}}]{planckNG:2015}%
  \BibitemOpen
\bibfield  {journal} {  }\bibfield  {author} {\bibinfo {author} {\bibfnamefont
  {P.~A.~R.}\ \bibnamefont {Ade}} \emph {et~al.} (\bibinfo {collaboration}
  {Planck}),\ }\href {\doibase 10.1051/0004-6361/201525836} {\bibfield
  {journal} {\bibinfo  {journal} {Astron. Astrophys.}\ }\textbf {\bibinfo
  {volume} {594}},\ \bibinfo {pages} {A17} (\bibinfo {year} {2016})},\ \Eprint
  {http://arxiv.org/abs/1502.01592} {arXiv:1502.01592 [astro-ph.CO]}
  \BibitemShut {NoStop}%
\bibitem [{\citenamefont {Shiraishi}\ \emph {et~al.}(2015)\citenamefont
  {Shiraishi}, \citenamefont {Liguori},\ and\ \citenamefont
  {Fergusson}}]{shiraishi/liguori/fergusson:2015}%
  \BibitemOpen
  \bibfield  {author} {\bibinfo {author} {\bibfnamefont {M.}~\bibnamefont
  {Shiraishi}}, \bibinfo {author} {\bibfnamefont {M.}~\bibnamefont {Liguori}},
  \ and\ \bibinfo {author} {\bibfnamefont {J.~R.}\ \bibnamefont {Fergusson}},\
  }\href {\doibase 10.1088/1475-7516/2015/01/007} {\bibfield  {journal}
  {\bibinfo  {journal} {JCAP}\ }\textbf {\bibinfo {volume} {1501}},\ \bibinfo
  {pages} {007} (\bibinfo {year} {2015})},\ \Eprint
  {http://arxiv.org/abs/1409.0265} {arXiv:1409.0265 [astro-ph.CO]} \BibitemShut
  {NoStop}%
\bibitem [{\citenamefont {Kamionkowski}\ and\ \citenamefont
  {Souradeep}(2011)}]{kamionkowski/souradeep:2011}%
  \BibitemOpen
  \bibfield  {author} {\bibinfo {author} {\bibfnamefont {M.}~\bibnamefont
  {Kamionkowski}}\ and\ \bibinfo {author} {\bibfnamefont {T.}~\bibnamefont
  {Souradeep}},\ }\href {\doibase 10.1103/PhysRevD.83.027301} {\bibfield
  {journal} {\bibinfo  {journal} {Phys. Rev.}\ }\textbf {\bibinfo {volume}
  {D83}},\ \bibinfo {pages} {027301} (\bibinfo {year} {2011})},\ \Eprint
  {http://arxiv.org/abs/1010.4304} {arXiv:1010.4304 [astro-ph.CO]} \BibitemShut
  {NoStop}%
\bibitem [{\citenamefont {Shiraishi}\ \emph {et~al.}(2013)\citenamefont
  {Shiraishi}, \citenamefont {Ricciardone},\ and\ \citenamefont
  {Saga}}]{shiraishi/ricciardone/saga:2013}%
  \BibitemOpen
  \bibfield  {author} {\bibinfo {author} {\bibfnamefont {M.}~\bibnamefont
  {Shiraishi}}, \bibinfo {author} {\bibfnamefont {A.}~\bibnamefont
  {Ricciardone}}, \ and\ \bibinfo {author} {\bibfnamefont {S.}~\bibnamefont
  {Saga}},\ }\href {\doibase 10.1088/1475-7516/2013/11/051} {\bibfield
  {journal} {\bibinfo  {journal} {JCAP}\ }\textbf {\bibinfo {volume} {1311}},\
  \bibinfo {pages} {051} (\bibinfo {year} {2013})},\ \Eprint
  {http://arxiv.org/abs/1308.6769} {arXiv:1308.6769 [astro-ph.CO]} \BibitemShut
  {NoStop}%
\bibitem [{\citenamefont {Matsumura}\ \emph {et~al.}(2014)\citenamefont
  {Matsumura} \emph {et~al.}}]{matsumura/etal:2013}%
  \BibitemOpen
  \bibfield  {author} {\bibinfo {author} {\bibfnamefont {T.}~\bibnamefont
  {Matsumura}} \emph {et~al.},\ }\href@noop {} {\bibfield  {journal} {\bibinfo
  {journal} {J. Low. Temp. Phys.}\ }\textbf {\bibinfo {volume} {176}},\
  \bibinfo {pages} {733} (\bibinfo {year} {2014})},\ \Eprint
  {http://arxiv.org/abs/1311.2847} {arXiv:1311.2847 [astro-ph.IM]} \BibitemShut
  {NoStop}%
\bibitem [{\citenamefont {Shiraishi}\ \emph {et~al.}(2016)\citenamefont
  {Shiraishi}, \citenamefont {Hikage}, \citenamefont {Namba}, \citenamefont
  {Namikawa},\ and\ \citenamefont {Hazumi}}]{shiraishi/etal:2016}%
  \BibitemOpen
  \bibfield  {author} {\bibinfo {author} {\bibfnamefont {M.}~\bibnamefont
  {Shiraishi}}, \bibinfo {author} {\bibfnamefont {C.}~\bibnamefont {Hikage}},
  \bibinfo {author} {\bibfnamefont {R.}~\bibnamefont {Namba}}, \bibinfo
  {author} {\bibfnamefont {T.}~\bibnamefont {Namikawa}}, \ and\ \bibinfo
  {author} {\bibfnamefont {M.}~\bibnamefont {Hazumi}},\ }\href {\doibase
  10.1103/PhysRevD.94.043506} {\bibfield  {journal} {\bibinfo  {journal} {Phys.
  Rev.}\ }\textbf {\bibinfo {volume} {D94}},\ \bibinfo {pages} {043506}
  (\bibinfo {year} {2016})},\ \Eprint {http://arxiv.org/abs/1606.06082}
  {arXiv:1606.06082 [astro-ph.CO]} \BibitemShut {NoStop}%
\end{thebibliography}%
\end{document}